\documentclass[preprint,showpacs,amsmath,amssymb,floatfix]{revtex4}
\usepackage{amsfonts}
\usepackage{amssymb}
\usepackage{amsmath}
\usepackage{epsfig}
\usepackage{hyperref}
\usepackage{bm}
\begin{document}
\title{Lift force on an asymmetrical obstacle immersed in a dilute granular flow}
\author{Fabricio Q. Potiguar}
\affiliation{Departamento de F\'\i sica, ICEN, Av. Augusto Correa, 1, Guam\'a, 
66075-110, Bel\'em, Par\'a, Brazil}
\begin{abstract}
This paper investigates the lift force exerted on an elliptical obstacle 
immersed in a granular flow through analytical calculations and computer 
simulations. 
The results are shown as a function of the obstacle size, orientation with 
respect to the flow direction (tilt angle), the restitution coefficient and 
ellipse eccentricity. The theoretical argument, based on the force 
exerted on the obstacle due to inelastic, frictionless collisions of a very 
dilute flow, captures the qualitative features of the lift, but fails to 
reproduce the data quantitatively. The reason behind this disagreement is that 
the dilute flow assumption on which this 
argument is built breaks down as a granular shock wave forms in front of the 
obstacle. More specifically, the shock wave change the grains impact 
velocity at the obstacle, decreasing the overall net lift obtained from a very 
dilute flow.
\end{abstract}
\maketitle

\section{Introduction}
Granular matter is a generic name given to a system composed of macroscopic, 
athermal particles that have mutual repulsive, dissipative interactions 
\cite{Jaeger96}. It is an intensely studied field in the Physics community 
given the several distinct behaviors shown by such systems as a consequence of 
different external conditions imposed on them.

One of such conditions is that which imposes a flow of particles, named 
granular flow \cite{Wieghardt75,Campbell90,Gold03,Midi04}. Within the several 
granular flow 
examples, the flow around immersed obstacles has received some attention 
lately \cite{Amar01,Rericha02,Boudet08}. One of the objectives of such 
investigations is to measure the force in the obstacle due to interactions 
with the flowing grains, the so called granular drag \cite{Albert99,Chehata03,Buchholtz98,Wass03,Ciamarra04,Soller06,Ding10}, analogously to the viscous flow 
force on an obstacle.

On the other hand, one knows that a viscous fluid 
flow around an obstacle produces an additional force called lift, 
which is perpendicular to the flow. Given the analogy between a viscous 
flow and the granular flow, a lift force should exist in 
an obstacle immersed in a granular flow under suitable conditions. However, 
most investigations focus only on the drag, while the lift studies are 
restricted only to a few experiments and simulations \cite{Soller06,Ding10}.

Soller and Koehler \cite{Soller06} showed that the lift on a rotating vane 
inserted in a granular packing scaled with geometric parameters of the 
system, such as an effective aspect ratio and immersion depth. The word 
effective means that the referred quantity, say the immersion depth, is 
considered taking into account the finite grain size. Ding {\em et al}. 
\cite{Ding10}, dragging an intruder 
at constant velocity through a granular packing, showed that a lift force is 
induced on the intruder and it depends on its geometry. Common to both 
investigations is the fact that the lift arises as a consequence of the 
hydrostatic nature of the stress in granular packings.

Given that these experiments were carried on such hydrostatic stress 
systems, variations of this condition might reveal different aspects of 
this force. For instance, is there a lift force in intruders immersed in flows 
where the stresses are not hydrostatic? If so, in what conditions this force 
arises? 

This paper is aimed at studying the lift force on an obstacle 
due to a dilute granular flow through numerical simulations. The approach will 
be identical to the one used in probing the drag on a cylinder due to 
a dilute granular flow \cite{Wass03}. The argument drawn here to obtain an 
analytical expression for the lift shows that there is no net lift in such 
dilute conditions on a circular obstacle. 
Hence, the obstacle chosen here is an elliptical one. The dependence of this 
force on the flow parameters will be obtained and studied numerically. 

Section \ref{sec1} is reserved for developing the theoretical argument and 
presenting its predictions. In section \ref{sim}, the simulation 
is described along with the numerical results for the lift. Section 
\ref{analy} holds analyses regarding the argument 
given in the previous section in order to understand the 
discrepancies between the theoretical and the numerical results. Finally, 
section \ref{conclusions} has the conclusions.

\section{Theory\label{sec1}}
In this section, the theoretical argument leading to an expression for the 
lift force on the ellipse is developed. Also, some predictions are shown in 
order to be compared to the numerical results in the next section.

\subsection{Argument development\label{sec11}}
In \cite{Wass03}, the line of thought that led to an expression for the drag 
force on the circular obstacle was based on 
the dissipative collisions among grains and obstacle. 
The same line of thought can be drawn here in order to obtain an expression 
for the lift force on the ellipse due to collisions with the incoming stream.

\begin{figure}[h]
\rotatebox{0}{\epsfig{file=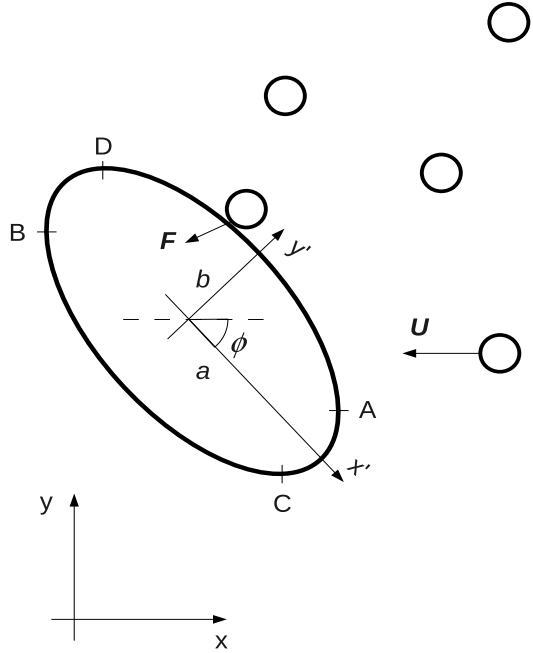,width=8.0cm,height=10.0cm}}
\caption{Ellipse's local frame of reference (primed system), rotated an 
angle $\phi$ about its center. The major and minor half-axes are given by $a$ 
and $b$, respectively. 
The force ${\bf F}$ is the total (normal plus tangential) force between a disk 
and the obstacle. See text for the meaning of the four points $A$, $B$, $C$ 
and $D$.
\label{local_ellipse_frame}}
\end{figure}

First of all, all disks are assumed to have only the horizontal velocity 
component, $U$. Therefore, for the ellipse given in fig. 
\ref{local_ellipse_frame}, which is located at $(x_E$,$y_E)$, a collision will 
occur only if the impact parameter 
(the vertical distance between a disk's 
center and the horizontal line through the ellipse's center) is in the range 
$\left[y_C+\frac{d}{2}\right.$,$\left.y_D+\frac{d}{2}\right]$. This is so 
because $C$ and $D$ are, respectively, the lowest and the highest points in the 
ellipse, in the same way as $A$ and $B$ are the leftmost and the rightmost 
ones. Any collisions that happen in the segment $\overline{AD}$ 
will exert a downward lift, while collisions that occur in the 
$\overline{AC}$ segment will produce an upward lift (from now on, the 
$\overline{AD}$ segment, and all other segments in the text, will be referred 
to only as $\overline{AD}$). Moreover, since the 
lengths of both segments are, in general, unequal, the longest of them, 
in this case $\overline{AD}$, will suffer more collisions than the other, 
which implies a net, negative, lift force. 
It is this mechanism that prevents any net lift to take place on a circular 
obstacle under these conditions (in general, any body that is symmetrical 
with respect to the flow direction, and is fully immersed in the flow, 
does not suffer a net lift). Therefore, the 
determination of the coordinates of these four points in the ellipse, $A$, 
$B$, $C$ and $D$, is very important.

In order to do this, one should notice that the tangent line to the ellipse is 
horizontal at $D$ and $C$ and vertical at $A$ and $B$. Therefore, starting 
from the tilted ellipse's equation in the global frame:
\begin{equation}
\label{ellipse_eq_glob_frame}
\left(a^2\sin^2\phi+b^2\cos^2\phi\right)X^2+\left(a^2\cos^2\phi+b^2\sin^2\phi\right)Y^2-\sin2\phi\left(a^2-b^2\right)XY=a^2b^2,
\end{equation}
where $X=x-x_E$ and $Y=y-y_E$, one can write for $D$ and $C$:
\begin{equation}
\label{x_DC}
x_{D,C}=x_E\pm\frac{\sin\phi\cos\phi(a^2-b^2)}{\sqrt{a^2\sin^2\phi+b^2\cos^2\phi}},
\end{equation}
and 
\begin{equation}
\label{y_DC}
y_{D,C}=y_E\pm\sqrt{a^2\sin^2\phi+b^2\cos^2\phi}.
\end{equation}
where the plus(minus) sign is for point $D$($C$). A similar calculation yields, 
for points $A$ and $B$:
\begin{equation}
\label{x_AB}
x_{B,A}=x_E\pm\sqrt{a^2\cos^2\phi+b^2\sin^2\phi},
\end{equation}
and
\begin{equation}
\label{y_AB}
y_{B,A}=y_E\pm\frac{\sin\phi\cos\phi(a^2-b^2)}{\sqrt{a^2\cos^2\phi+b^2\sin^2\phi}},
\end{equation}
where in both eqs. the plus(minus) sign is for point $B$($A$).  

The force due to a collision between a disk and the ellipse can be obtained by 
calculating the change in linear 
momentum of the disk. The total force is obtained by integrating the product 
of this individual momentum change and a suitable collision rate over the 
obstacle section facing the flow, i.e., integrating over $\overline{CD}$.

\begin{figure}[h]
\rotatebox{0}{\epsfig{file=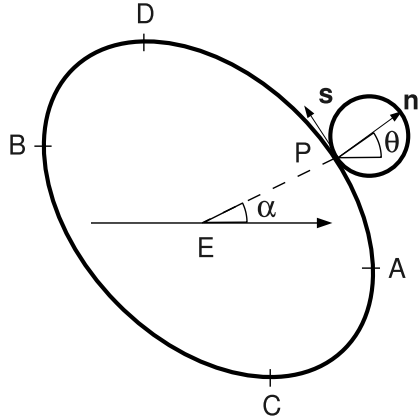,width=10.0cm,height=7.0cm}}
\caption{Geometry of the collision between a disk and the ellipse, centered at 
$E$. They touch each other at the point $P$ in the ellipse. $\theta$ and 
$\alpha$ are the impact and collision angles.
\label{coll_geo}}
\end{figure}

Suppose a disk hits the obstacle in $\overline{AD}$. Figure 
\ref{coll_geo} presents schematically the geometry of the collision (assumed 
frictionless).

The grain velocity is ${\bf v}_0=-U{\bf i}$, and the impact angle is $\theta$, 
which is the one between the direction ${\bf i}$ and the normal at the 
contact point. 
The post-collisional velocity is ${\bf v}=v_x{\bf i}+v_y{\bf j}$.  In order to 
obtain the final components as a function of $U$, one has two equations 
that relate the normal and tangential velocities before and after the 
collision: ${\bf v}\cdot{\bf\hat n}=-e{\bf v}_0\cdot{\bf\hat n}$ and 
${\bf v}\cdot{\bf\hat s}={\bf v}_0\cdot{\bf\hat s}$, where
${\bf\hat n}=\cos\theta{\bf i}+\sin\theta{\bf j}$, 
${\bf\hat s}=-\sin\theta{\bf i}+\cos\theta{\bf j}$ are the normal and 
tangential unit vectors at the collision point and $e$ is the normal 
restitution coefficient, assumed velocity independent.

From these considerations, the components of the disk velocity after the 
collision are 
$v_x=U(\sin^2\theta-e\cos^2\theta)$ and $v_y=\frac{1}{2}U(1+e)\sin2\theta$. 
Therefore, the vertical momentum change for a general collision is given by:
\begin{equation}
\label{py_change}
\Delta P_y(\theta)=mv_y=\frac{1}{2}mU(1+e)\sin2\theta.
\end{equation}
 
The number of disks that strike the ellipse in a time $dt$ is given by the 
number of particles within the area of the parallelogram with sides $Udt$ and 
$ds$, this last quantity being the arc length around the collision point. Given 
the geometry of the collision (figure \ref{coll_geo}), the area of this 
parallelogram is $dA=U\cos\theta dtds$. Hence:
\begin{equation}
\label{coll_freq}
dN=\frac{4\nu}{\pi d^2}dA=\frac{4\nu}{\pi d^2}U\cos\theta dtds,
\end{equation}
where $\nu$ is the area fraction of the incoming flow. The collision frequency 
is simply this number divided by $dt$. 

\begin{figure}[h]
\rotatebox{0}{\epsfig{file=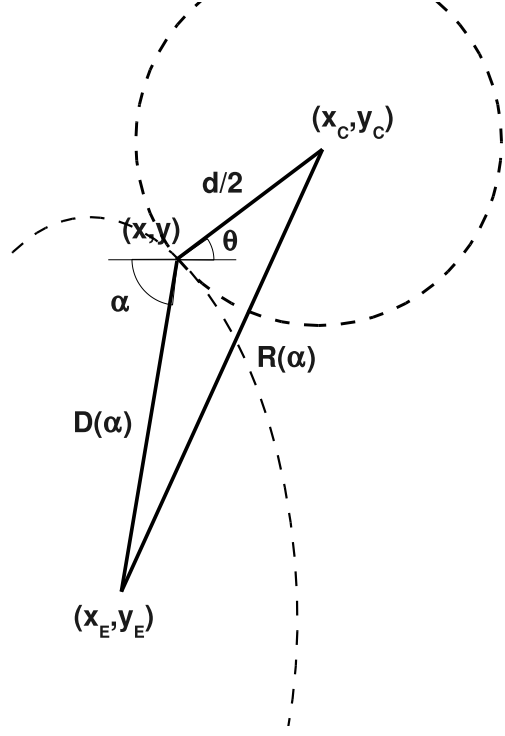,width=7.5cm,height=10.0cm}}
\caption{Triangle formed by the disk center, $(x_C,y_C)$, the ellipse center, 
$(x_E,y_E)$ and the collision point, $(x,y)$ (angles are exaggerated). The 
angle between $D(\alpha)$ and the normal to the disk center is 
$\pi-\alpha+\theta$. Dashed lines are sketches of the surfaces of the disk and 
the ellipse. \label{coll_tri}}
\end{figure}

The arc length $ds$ is given by the 
product of the collision radius, $R(\alpha)$, by the arc element, $d\alpha$. 
This product is readily evaluated for a circle. However, in the ellipse 
case, this is not so simple, since this radius varies with $\alpha$. Besides, 
the grain size introduces an additional complication for calculating 
$R(\alpha)$, as seen in fig. \ref{coll_tri}, which shows the triangle formed 
by the disk center, the ellipse center and the collision point. 
Hence, the ellipse is assumed much larger than the disks and the disk size 
only enters the expression for $R(\alpha)$ as a minor correction. With this 
assumption, the arc length is given by $ds=D(\alpha)d\alpha$. In parametric 
form, $x=a\cos\alpha$ and $y=b\sin\alpha$, this length reads:
\[
ds=\sqrt{\left(\frac{dx}{d\alpha}\right)^2+\left(\frac{dy}{d\alpha}\right)^2}d\alpha=(\sqrt{a^2\cos^2\alpha+b^2\sin^2\alpha})d\alpha.
\]

Finally, the effective collision radius is given by:
\begin{equation}
\label{radius_param}
D(\alpha)=(a^2\cos^2\alpha+b^2\sin^2\alpha)^{1/2}.
\end{equation}

Another difficulty to obtain the expression for the lift force is that 
the momentum change (i.e., the force) depends explicitly on $\theta$, while 
the collision frequency depends on $\alpha$. Therefore, any hope of integrating 
the lift force over all collisions should pass through obtaining the 
relationship between the angles $\theta$ and $\alpha$. This is done in the 
following paragraph.

Since $dx/dy$ gives the tangent of the angle between the normal line to a 
point in the ellipse and the $x$ axis, 
from the ellipse equation (\ref{ellipse_eq_glob_frame}), one has:
\[
\tan\theta=\frac{2C_2Y-C_3X}{C_3Y-2C_1X},
\]
where $C_1$, $C_2$ and $C_3$ are the coefficients of the terms $X^2$, $Y^2$ 
and $XY$ in eq. (\ref{ellipse_eq_glob_frame}). Since $\tan\alpha=Y/X$ and 
from eqs. (\ref{x_DC}), (\ref{y_DC}), (\ref{x_AB}) and (\ref{y_AB}), the 
relation between 
$\theta$ and $\alpha$ can be cast in terms of the coordinates of the points 
$D$ and $A$, since:
\[
\tan\alpha_D=\frac{y_D-y_E}{x_D-x_E}=
\frac{a^2\sin^2\phi+b^2\cos^2\phi}{(1/2)\sin2\phi(a^2-b^2)}=
\frac{2C_1}{C_3}
\]
and 
\[
\tan\alpha_A=\frac{y_A-y_E}{x_A-x_E}=
\frac{(1/2)\sin\phi\cos\phi(a^2-b^2)}{a^2\cos^2\phi+b^2\sin^2\phi}=
\frac{C_3}{2C_2}.
\]

Hence:
\begin{equation}
\label{theta_alpha_rel}
\tan\theta=\frac{1}{\tan\alpha_A}\frac{\tan\alpha_A-\tan\alpha}{\tan\alpha_D-\tan\alpha}.
\end{equation}

The lift force can now be evaluated as:
\[
L=-\int\Delta P_yd\dot{N}
\]
where $d\dot{N}=dN/dt$. By using eqs. (\ref{py_change}), (\ref{coll_freq}) and 
(\ref{radius_param}), the total lift on the ellipse due to collisions is given 
by:
\begin{equation}
\label{lift_AD}
L=-\frac{1}{2}\rho U^2(1+e)\int\limits_{C}^{D}\sin2\theta\cos\theta 
ds(\alpha),
\end{equation}
where $\rho=4m\nu/\pi d^2$ is the flow mass density and the arc length is to 
be calculated bearing in mind (\ref{theta_alpha_rel}). 

\subsection{Model results\label{sub22}}
The lift force scales with the obstacle size $a$, as expected, since it 
depends linearly on the arc length. Also, it is seen that, for more inelastic 
grains the lift is smaller, since more inelasticity implies less intense 
change of momentum, see eq. (\ref{py_change}). The dependence of the lift on 
the tilt angle is 
hidden in the integral over $\overline{CD}$, because the lengths of 
$\overline{AC}$ and $\overline{AD}$, which contribute forces with distinct 
signs, are unequal for a general $\phi$. 

\begin{figure}[h]
\rotatebox{0}{\epsfig{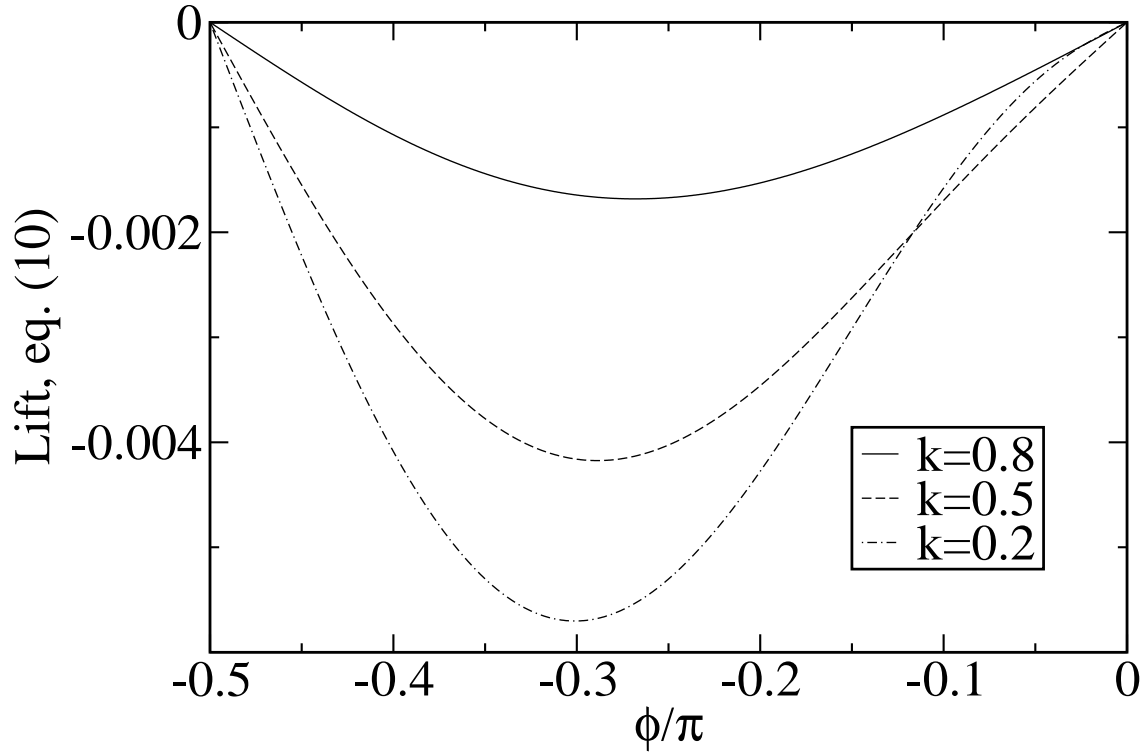}}
\caption{Eq. (\ref{lift_AD}) for distinct eccentricity, $k$, values. The 
  curves were obtained with $\nu=0.196$, $d=1$, $m=1$, $U=10$ and $e=0.952$.
\label{lift_theo_k}}
\end{figure}

Apart from the dependence of $L$ in these parameters, it also depends on the 
ellipse eccentricity, i.e., on the ratio of the ellipse's axes 
$0\leq k=b/a\leq1$. 
For $k=1$, the obstacle is a circle, and the lift vanishes by symmetry. In the 
other extreme value, $k=0$, which corresponds to a flat plate, the lift 
vanishes 
only for $\phi=0$ (horizontal plate) and $\phi=\pm\pi/2$ (vertical plate), 
which are the only symmetrical orientations with respect to an horizontal 
flow. In fig. \ref{lift_theo_k}, it is shown the predictions of eq. 
(\ref{lift_AD}) for distinct values of $k$ (the parameter values in plotting 
these curves were chosen to agree with those used in the simulations).

One can see that the maximum (absolute) value of $L$ increases when $k$ 
decreases. This happens because when $k\rightarrow0$, the points $D$ and $C$ 
merge with points $B$ and $A$, as can be inferred from eqs. (\ref{x_DC}), 
(\ref{y_DC}), (\ref{x_AB}), and (\ref{y_AB}). In this case, 
$\overline{AD}\rightarrow2a$ and $\overline{AC}\rightarrow0$. Since the lift is 
the difference of the contributions from $\overline{AD}$ and $\overline{AC}$, 
all collisions will give positive contributions to the net lift value, and it 
should reach its maximum for a particular $\phi$.

The objective of the simulations 
is to study the lift force on the obstacle as a function of four 
parameters that appear on eq. (\ref{lift_AD}), namely, the obstacle size, $a$, 
the restitution coefficient, $e$, the ellipse eccentricity, $k$, and the tilt 
angle, $\phi$.

\section{Simulation\label{sim}}
In this section, the simulation is detailed, along with the parameter values 
used in the computations, and the numerical results for the lift are shown. 

\subsection{Design\label{sub21}}
The system is composed of $N$ soft disks, with unity diameter $d$ and unity 
mass $m$, 
located in a workspace of lengths $L_X$ and $L_Y$ 
in the horizontal and vertical directions, respectively. They interact through 
normal and tangential forces, according to the model 
used in \cite{rapaport,Ciamarra04}. The normal force between two disks is:
\begin{equation}
\label{norm_force}
{\bf F}_{ij}^N={\bf f}_{ij}+{\bf f}_{ij}^d,
\end{equation}
where the first term is the conservative part, given by a simple harmonic 
spring force
\[
{\bf f}_{ij}=\kappa\left(\frac{d_i+d_j}{2}-r_{ij}\right){\bf{\hat r}}_{ij},
\]
where $\kappa$ is the spring constant, $d_i$ is the $i$-th disk diameter, 
$r_{ij}$ is the distance between the two disks and ${\bf{\hat r}}_{ij}$ is a 
unit vector along the normal between the disks' centers. The second term is a 
dissipative, velocity dependent force, given by
\[
{\bf f}_{ij}^d=-\gamma_d({\bf {\hat r}}_{ij}\cdot {\bf v}_{ij}){\bf{\hat r}}_{ij},
\]
where $\gamma_d$ is the normal damping coefficient and 
${\bf v}_{ij}={\bf v}_i-{\bf v}_j$ is the 
relative velocity between the contacting disks. The tangential 
force at the contact point is given by the following expression: 
\begin{equation}
\label{tang_force}
{\bf F}_{ij}^S=-\min\left(\gamma_sv_{ij}^S,\mu F_N\right){\bf{\hat v}}_{ij}^S,
\end{equation}
where $\gamma_s$ is the sliding friction constant and $\mu$, the static 
friction coefficient, while ${\bf{\hat v}}_{ij}^S$ is the unit vector along the 
direction of the relative velocity at the contact point. This vector is 
calculated as: 
\[
{\bf v}_{ij}^S={\bf v}_{ij}-({\bf {\hat r}}_{ij}\cdot {\bf v}_{ij}){\bf{\hat r}}_{ij}-
\left(\frac{{\bm{\omega}}_i+{\bm\omega}_j}{2}\right)\times{\bf r}_{ij},
\]
where ${\bm\omega}_i$ is the $i$-th disk angular velocity and 
$v_{ij}^S=\left|{\bf v}_{ij}^S\right|$. The total contact force 
${\bf F}_{ij}={\bf F}_{ij}^N+{\bf F}_{ij}^S$ vanishes if the disks are not in 
contact, i.e., if $\frac{d_i+d_j}{2}>r_{ij}$.

The values of the elastic parameters used were: $\kappa=50000$, $\gamma_d=10$ 
and $\gamma_d=100$, $\gamma_s=0.1$ and $\mu=1$. The restitution coefficient 
\cite{Silbert01} for these parameters were $e=0.952$, for $\gamma_d=10$, and 
$e=0.608$, for $\gamma_d=100$. 
The ellipse's major and minor half-axes are $a$ and $b=ka$, respectively. 
 The parameter $a$ has the values $5$, $10$, $20$, $30$ and $40$, with $k=0.8$. 
For the studies of the dependence on $k$, size $a=10$ was used, with 
$k=0.80$, $0.50$ and $0.20$. All these cases were studied for both $e$ values. 
For the two smaller obstacle sizes, results were obtained with a 
$N=2500$ and $L_X=L_Y=100$ packing, 
while for the others, a $N=10000$ and $L_X=L_Y=200$ packing was used. In both 
cases, the packing fraction was about $0.200$. The ellipse is located in 
$x_E=0.5L_X$ and $y_E=0.5L_Y$, and its major half-axis is tilted related to the 
horizontal axis (flow direction) by $\phi$. The values of the tilt angle were 
in the range $[-\pi/2$,$0]$, divided in $\pi/20$ increments, which gives a 
total of $11$ distinct $\phi$ values.

At the beginning, all disks are randomly generated without overlap among them 
and the ellipse. 
At this stage, the obstacle is modeled as a circle with diameter $2a$ since 
this facilitates the overlap check. All disks have initial 
velocity ${\bf v}_0=-U{\bf i}$, where ${\bf i}$ is the unit vector in 
the horizontal (flow) direction and $U=10$.

The system has periodic boundaries perpendicular to the flow direction, while, 
along the flow, the conditions are the same as in \cite{Wass03}: whenever a 
disk 
leaves the system through the right boundary, it is placed at the left 
one in a random vertical position. In fact, the code searches for a position 
where the incoming grain overlaps with no other disk, which is fairly easy, 
giving the low density that is used here. Its velocity is set as the incoming 
flow velocity ${\bf v}_0$ plus a small random vertical component chosen 
uniformly in the interval $[-fU,fU]$, where $f=0.10$ (results obtained with 
$f=0$ differ little from those shown here). 


The interactions between the disks and the ellipse are the same as those 
given above for two disks. The main problem is to resolve particle-ellipse 
contacts. In order to do this, all disks are checked for 
overlap with a circle of diameter $2a$. If one disk overlaps with the circle, 
it is mapped in the ellipse's local frame of reference (ELFR). Then, the 
contact point coordinates, in the ELFR, are calculated using the algorithm 
proposed in \cite{Dziu01}.
The direction of the normal force 
is along the line joining the disk center and the contact point, while the 
tangential force is perpendicular to this direction.

Lengths, forces and time are given in units of $d$, $\kappa d$, and 
$\sqrt{m/\kappa}$, respectively. Equations of motion are integrated with a 
leapfrog scheme \cite{rapaport}, with a time step of $0.001$. Each simulation 
is performed during $10^5$ molecular dynamics (MD) cycles for thermalization 
and $10^6$ MD cycles for measurements. All results are averaged over $20$ and 
$10$ independent runs for the $N=2500$ and $N=10000$ packings, respectively.
\begin{figure}[h]
\rotatebox{0}{\epsfig{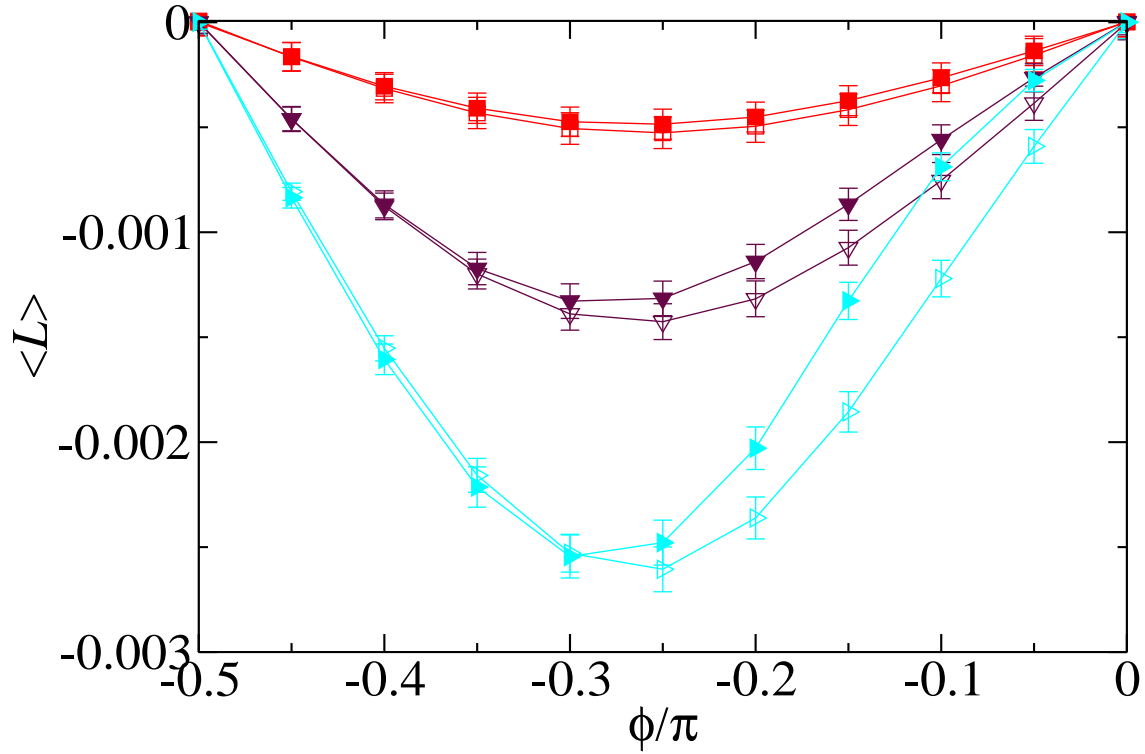}}
\caption{(Color online) Lift as a function of eccentricity, $k$, restitution 
coefficient and tilt angle, $\phi$. Squares represent $k=0.80$ 
results, inverted triangles, $k=0.50$ and left triangles, $k=0.20$. Open and 
filled symbols are for $e=0.952$ and $0.608$. All results are for $a=10$. 
Lines are only guides to the eyes.
\label{lift_k}}
\end{figure}
The main interest is to measure the force exerted on the obstacle by the 
stream of grains due to the collisions among them. Therefore, the components of 
this force, drag and lift, are measured at all cycles after thermalization. At 
regular intervals, the forces are recorded as averages over this period. This 
accumulation interval is $1000$ MD cycles long (results were also obtained with 
$10000$ MD cycles long accumulation periods and do not differ appreciably from those 
reported here). The flow velocity and particle number density fields, 
${\bf v}(x,y)$ and $\rho(x,y)$, were measured as follows: 
the workspace is divided in square bins of side $d$. At each cycle, all 
particles are mapped in an appropriate bin and its velocity components are 
added to the respective field element. Similarly, angular profiles related to 
the contact angle between the grains and the ellipse, the collision angle 
$\alpha$, were measured. They are the collision number and the velocity 
components. Each time there is a disk-ellipse contact, the angle formed by the 
line joining the collision point and the ellipse center with the flow 
direction is calculated. Then, the above mentioned quantities are added 
to the appropriate profile bins. The field and the profiles are presented as 
averages over cycles and runs.

\subsection{Results\label{res}}
In fig. \ref{lift_k}, the numerical results for the lift as a function of the 
tilt angle and the eccentricity are shown. 

 Comparing figs. \ref{lift_theo_k} and \ref{lift_k}, it is clear that eq. 
(\ref{lift_AD}) captures the qualitative features of the lift force, even 
though the correspondence becomes weak for $k=0.20$ and $e=0.952$. Also, the 
results agree with the prediction that the lift should be higher for less 
inelastic flows. Finally, the theoretical result overestimates the numerical 
ones by, roughly, a factor of $2$, for these cases.

\begin{figure}[h]
\rotatebox{0}{\epsfig{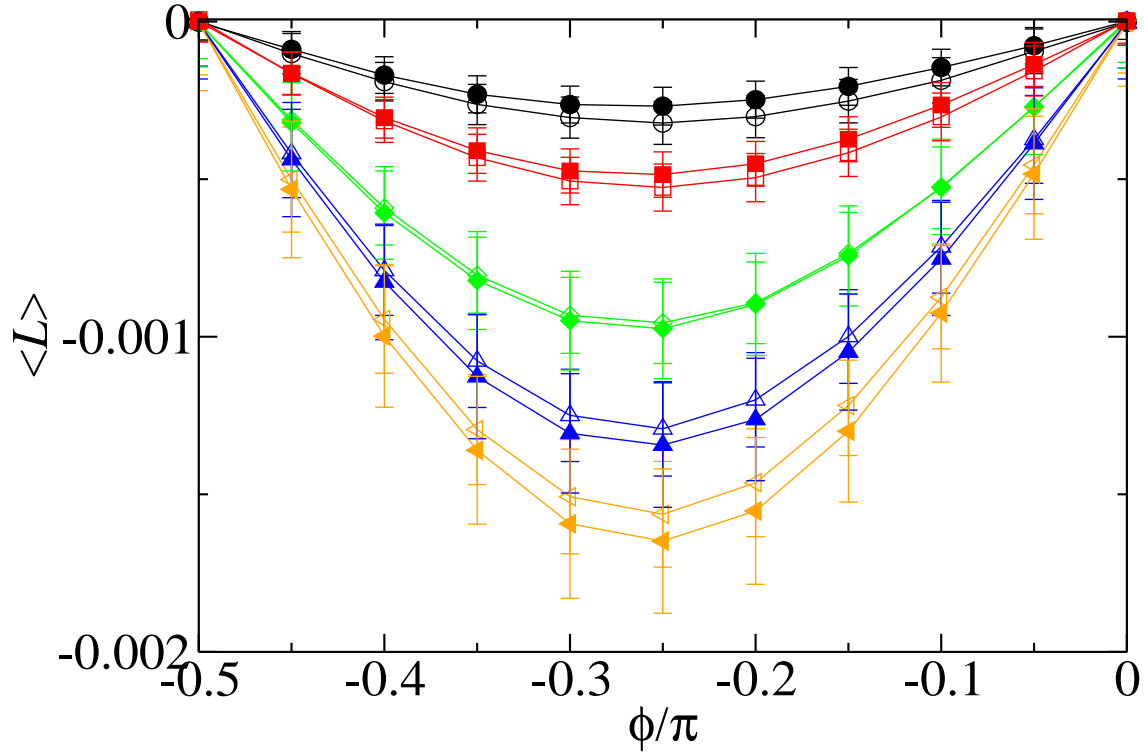}}
\caption{(Color online) Lift force as a function of the tilt angle, 
obstacle size, $a$, and restitution coefficient $e$. 
Distinct symbols are $a=5$ (black circles), $a=10$ (red squares), $a=20$ 
(green diamonds), $a=30$ (blue triangles) and $a=40$ (orange left triangles). 
All are for $k=0.80$. Open and filled symbols are results for $e=0.952$ and 
$e=0.608$, respectively. Lines are only guides to the eyes.
\label{lift_phi}}
\end{figure}

In fig. \ref{lift_phi}, the numerical results for the lift force as a function 
of the tilt angle and obstacle size are shown. As expected, the lift increases 
with the obstacle size. However, the lift does not grow linearly with $a$. 
Plotting these data against the obstacle size shows that 
$\left<L\right>\sim a^{0.78(2)}$. Since the results are not conclusive, it 
suffices to acknowledge that more simulations are needed in order to obtain 
a more reliable scaling with the obstacle size. Also, 
this figure shows that the lift force for $e=0.608$ packings grows faster with 
$a$ than for those with $e=0.952$. Finally, the numerical results, as seen in 
fig. \ref{lift_k}, also are smaller than their theoretical counterparts 
(for $a=40$, the factor is about $4$). This shows that the difference between 
model and theory depends only a little on the tilt angle, while the obstacle 
size, inelasticity and eccentricity are the factors that accept the most 
on this difference.

The reasons behind the failure to reproduce the numerical results will be 
analyzed in the next section, where the assumption that led to eq. 
(\ref{lift_AD}) will be reviewed in detail.

\section{Analyses\label{analy}}
\begin{figure}[h]
  \rotatebox{0}{\epsfig{file=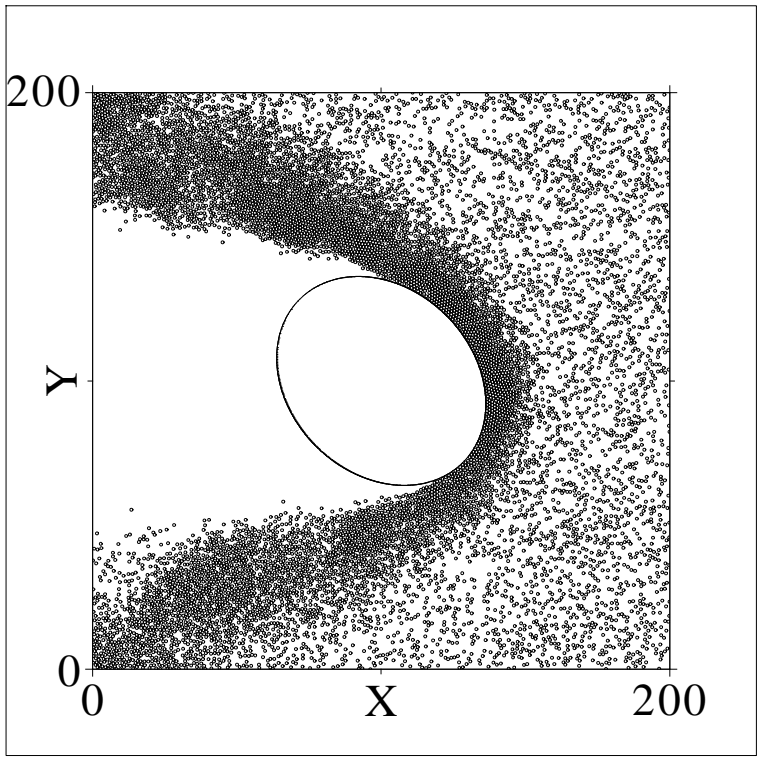,width=10.0cm,height=10.0cm}}
\caption{System configuration. Parameters: $a=40$, $k=0.80$, $e=0.952$ and 
$\phi=-\pi/4$.
\label{config_4_6}}
\end{figure}

There are three basic hypotheses on which the argument for the lift force in 
section \ref{sec11} was built. The first one is that particle-obstacle 
interactions were frictionless, despite the force model, eq. 
(\ref{tang_force}), is not. Second, the arc length expression, 
$ds=D(\alpha)d\alpha$, 
does not take into account the fact that the grains have a finite diameter $d$. 
Finally, the collision frequency expression (\ref{coll_freq}) was 
obtained under the hypothesis that only one particle at a time hits the 
obstacle, i.e., a disk-ellipse collision does not affect the next 
collision, a condition met only in very dilute or in ideal gas flows (which is 
not the case here). Therefore, its 
applicability here is clearly questionable, as inferred from the system 
configuration shown in fig. \ref{config_4_6}. 

\begin{figure}[h]
\centering
\begin{tabular}{c}
\rotatebox{0}{\epsfig{file=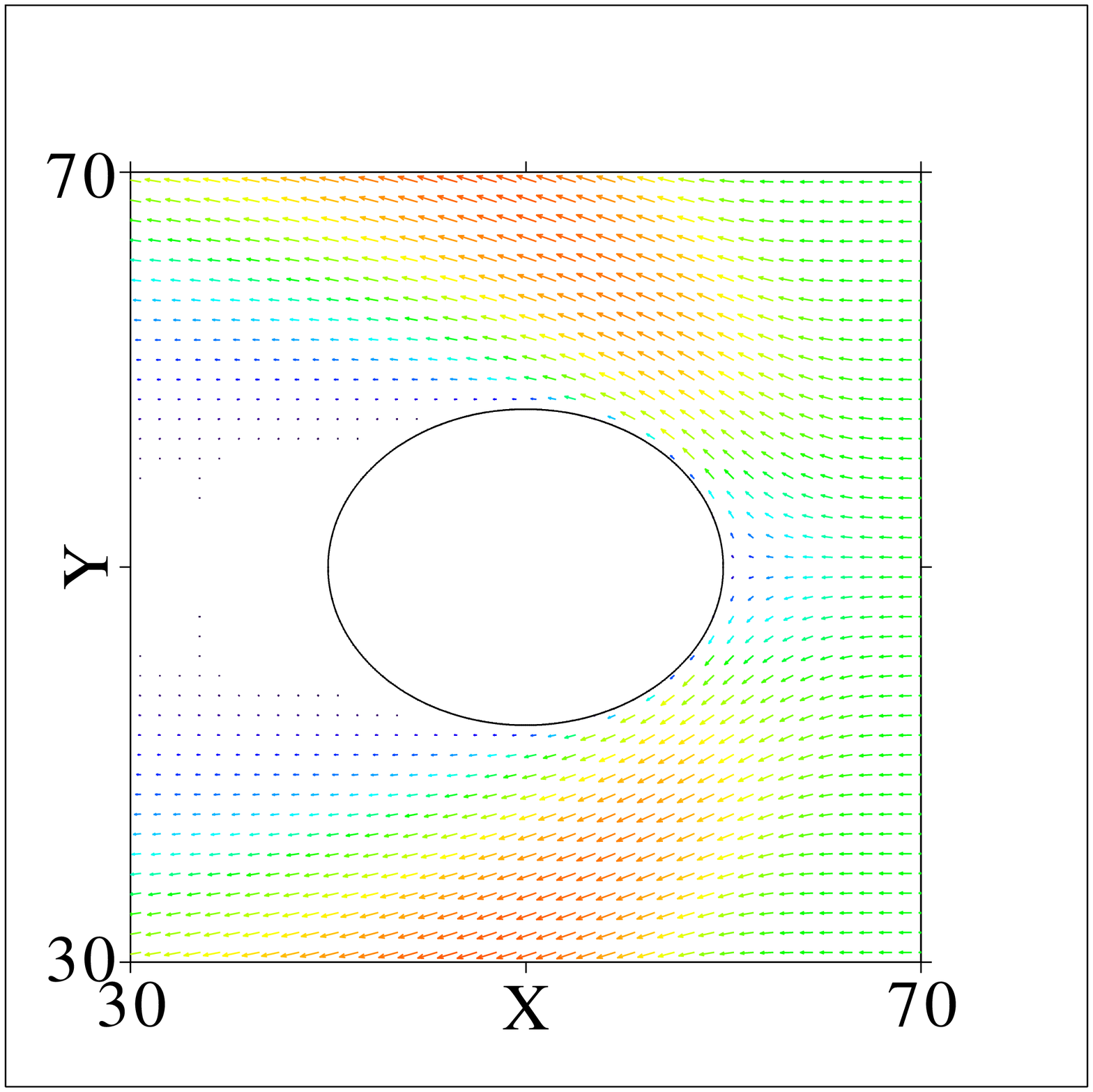,width=10.0cm,height=10.0cm}}\\
\rotatebox{0}{\epsfig{file=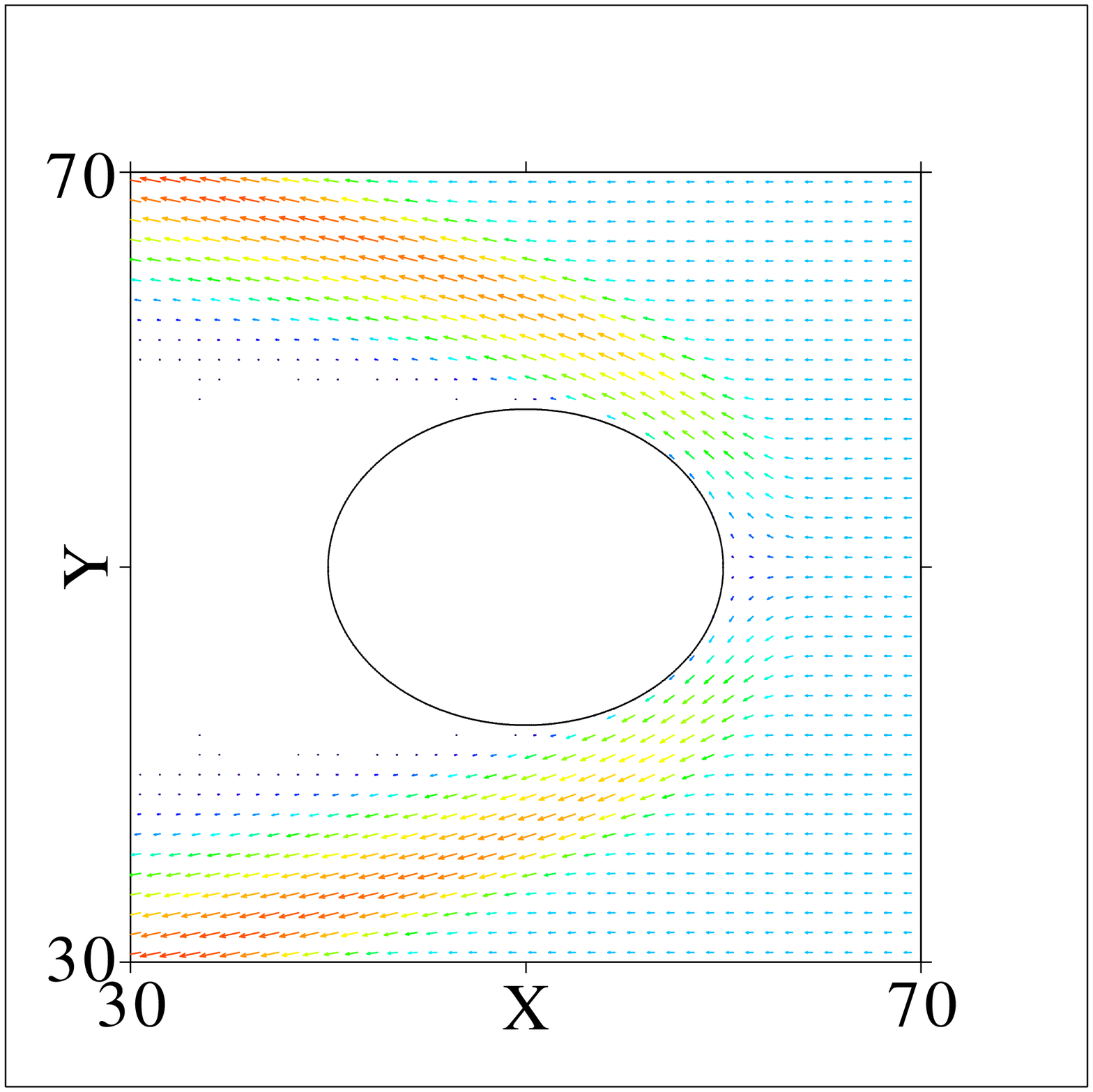,width=10.0cm,height=10.0cm}}\\
\end{tabular}
\caption{(Color online) Velocity fields detailed around the obstacle for 
$a=10$, $k=0.80$, $\phi=0$, $e=0.952$ (top) and $e=0.608$ (bottom).
\label{vel_fields}}
\end{figure}

There is a dense region that begins above point $A$ and forms a separation 
boundary in front of the ellipse. This is the typical granular shock wave 
\cite{Amar01,Rericha02,Boudet08,Buchholtz98,Wass03} that forms around bodies 
immersed in fast 
granular flows. Clearly, the dilute flow assumption is not valid. 
Figure \ref{vel_fields} shows the clear signature of this 
structure in the velocity field. 

Also, this picture shows that the two branches of the shock wave meet behind 
the obstacle. This fact could, at first sight, invalidates the lift results 
because this clearly does not allow the use of periodic boundaries 
perpendicular to the stream. This is not the case, however, as seen in the 
results for the lift obtained from simulations that used rigid walls in the $y$ 
direction. These results agree with those shown in figs. \ref{lift_k} and 
\ref{lift_phi} within numerical error. Nevertheless, periodic boundaries in a 
particular direction should only be used when the system is independent in that 
direction.


Before discussing the influence of the shock wave in the results, a few brief 
comments will be made regarding the first two assumptions in the 
theoretical argument. 

To take into account the grain size in computing the collision radius is to 
consider a larger (effective) obstacle facing the flow, which would increase 
the theoretical prediction for the lift. Since this quantity is already larger 
than the numerical results, it was safe to ignore it in the calculations.

The influence of the friction force on the lift can be inferred from the 
velocity field results, fig. \ref{vel_fields}. As seen in this figure, 
particles follow tangential trajectories along the ellipse
\cite{Amar01,Rericha02,Boudet08}, which means that they slide along the 
obstacle. Since friction is a tangential force, one concludes that 
a grain sliding along $\overline{AD}$ exerts an upward lift. By symmetry, a 
disk exerts a downward lift while sliding along $\overline{AC}$. Therefore, 
the net effect of friction is to decrease the lift exerted by collisions in 
each segment. The combination of both effects 
might decrease or increase the net lift. The numerical results for the lift 
rising from friction show that it is opposite to the one resulting from the 
normal force. In other words, friction decreases the net lift. The effect is 
small, though, because the sliding friction is also small, $\gamma_s=0.1$. In 
most simulations of granular 
matter, regardless of the tangential force model, the sliding friction is only 
mildly lower than the normal interaction parameter. Here, a much lower value 
was used, which could greatly affect the results. Simulations performed with a 
distinct set of parameters, $\gamma_s=5000$ and $\mu=0.1$, values which are 
more common to general simulations of granular matter, show that the 
results do not change significantly, and are still qualitatively the same as 
those in figs. \ref{lift_k} and \ref{lift_phi}. In fact, the lift data are 
reduced only by $5\%$ to $10\%$. This happens due to 
the fact that the velocities involved in the simulations are high enough for 
the tangential force to reach the Coulomb static friction criterion, which 
caps its maximum value and limits its influence on the lift. 
This conclusion is supported by the fact that the fraction of all 
disk-obstacle contacts, for $\gamma_s=5000$, that reach the failure condition 
is about $99.5\%$, with only a very small $\phi$ dependence.

As stated earlier, the dilute flow assumption does not hold. A rigorous 
calculation of the force in the obstacle, in which all 
dense packing effects are taken into account, is not a simple matter. In 
\cite{Deng10} an attempt was made in this direction. 
The formation of the shock wave produces a dense region that shields the 
obstacle from the incoming particles. Since the interactions are dissipative, 
particles should reach the obstacle with reduced velocities. This clearly 
reduces the force exerted by the disks.
Also, given the large density within the shock wave, the collision rate might 
be affected. Finally, as seen from 
the velocity field data, fig. \ref{vel_fields}, particles should hit the 
obstacle, on average, with a non vanishing vertical velocity component, i.e., 
particles suffer 
oblique inelastic collisions, which could affect the force on the obstacle. 
Both effects will 
be discussed in more detail in the following two 
subsections. This discussion will by made only qualitatively, since more 
simulations are needed to determine the amount of influence each of them has 
on the results.

\subsection{Obstacle shielding\label{sub31}}
One of the consequences of the existence of the shock wave is that the obstacle 
is shielded from the flow, in a way that the grains hit it with lower 
horizontal velocity than the upflow value. This fact alone indicates that the 
numerical net lift should be lower than its theoretical prediction. Aside from 
reducing the incoming flow velocity, the shock wave forms a dense region 
around the obstacle through which grains should go through in order to reach 
it. This could affect the collision rate and, in turn, the net lift. Both 
effects must play a role in order to explain the faster growth of the lift 
force with the obstacle size for stronger inelasticity flow compared to that 
of lower inelasticity one. 

\begin{figure}[h]
\rotatebox{0}{\epsfig{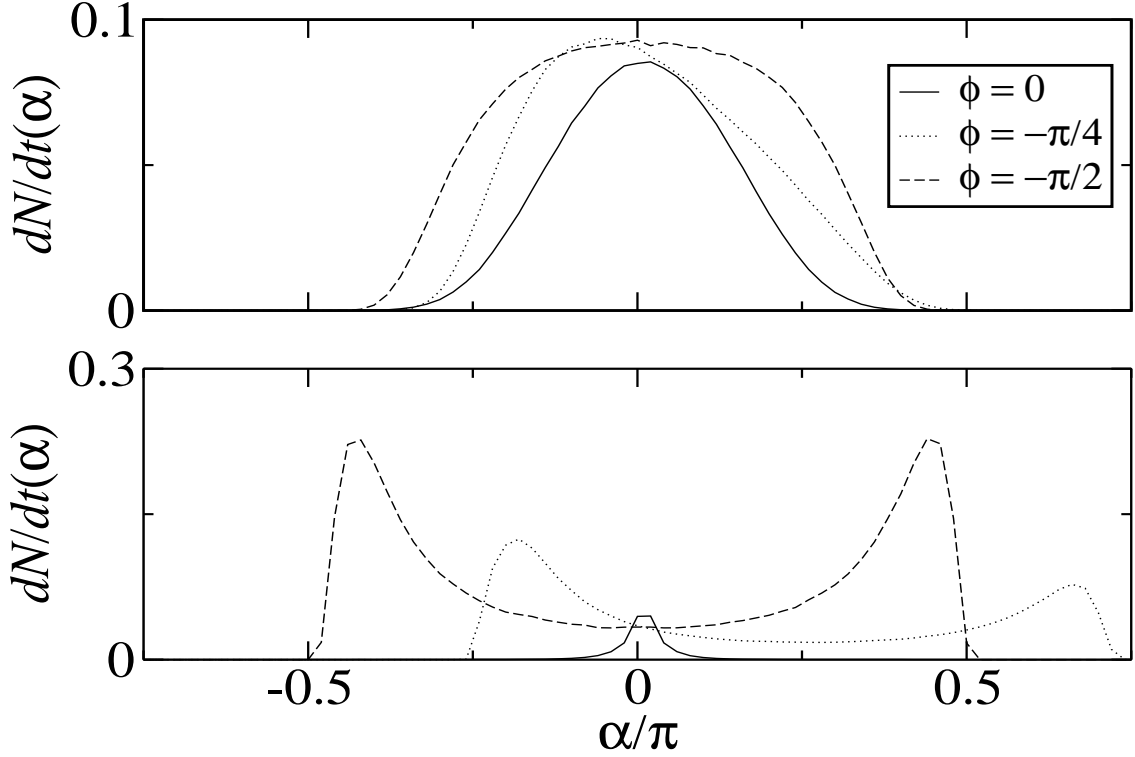}}
\caption{Angular collision profiles, for $a=10$, $e=0.608$, $k=0.80$ 
(upper row) and $k=0.20$ (lower row), for three distinct tilt angles (shown in 
the legend box). The horizontal scales are the same for both plots.
\label{coll_ang_prof_2}}
\end{figure}

The results for the collision profiles show that, for $e=0.608$, 
the shock wave is more localized around point $A$ (where, for the $k=0.80$ 
obstacles the peak of the profiles is located) and that the overall collision 
number is greatly increased compared to the number when $e=0.952$. This is a 
consequence of the 
aggregation that grains suffer due to the inelastic interactions and is 
common to all cases studied. This fact 
implies that the force on the obstacle should increase due to this increase in 
the collision rate and could compensate, in part, for the decrease in the force 
due to the reduced incoming velocity. 

\begin{figure}[h]
\rotatebox{0}{\epsfig{file=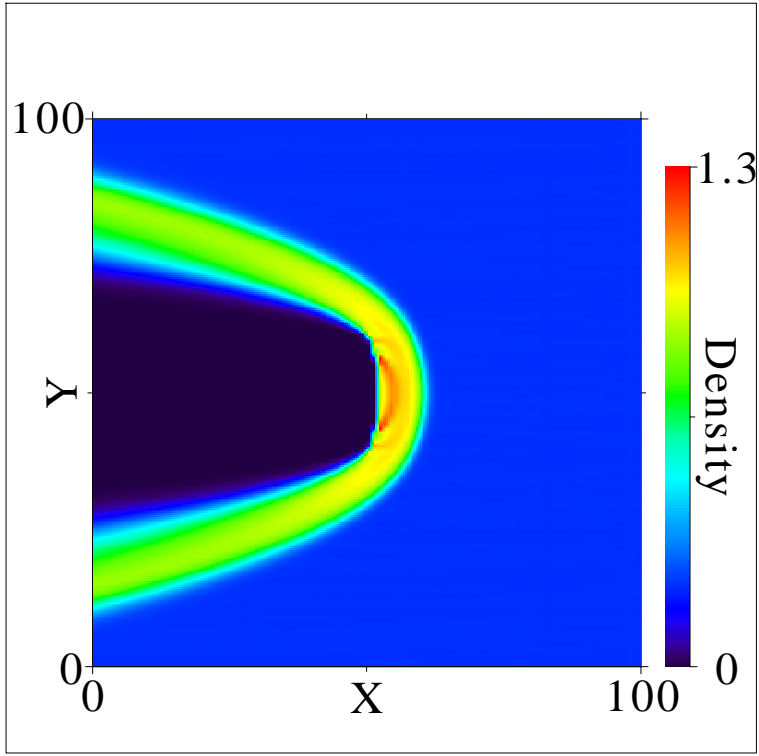,width=10.0cm,height=10.0cm}}
\caption{(Color online) Density field for $a=10$, $e=0.608$, $k=0.80$ and 
$\phi=-\pi/2$. The gap is the yellow (white) region between the red 
(light gray) arch and the ellipse (not shown).
\label{rho_f}}
\end{figure}

The collision profiles are changed, in a very different way, for obstacles 
with distinct eccentricities, $k$. Fig. \ref{coll_ang_prof_2} has results that 
illustrate this effect. The most striking feature of these curves is that the 
profiles for $k=0.20$ 
develop two peaks, instead of the more familiar one around point $A$, which is 
still present, but it moves away from this point as the ellipse is oriented 
vertically. The other one develops around point $D$ and is seen at tilt 
angles as high as $\phi=-\pi/10$. Such features are also seen in the $k=0.50$ 
results, although with smaller peaks. These results can be explained by the 
fact that in some cases, for stream velocities below a critical value, there 
is the appearance of a gap, 
filled with a hot granular gas, between the obstacle and the shock wave 
\cite{Buchholtz98}. In the present case, the appearance of the gap is a 
function of $k$ alone, since the stream velocity is the same for all 
simulations. Figure \ref{rho_f} has a result for the density field that 
indicates the presence of the gap. Notice the denser arc formed right in front 
of the ellipse, the region limited by this arc and the obstacle is the gap.

\begin{figure}[h]
\centering
\begin{tabular}{c}
\rotatebox{0}{\epsfig{file=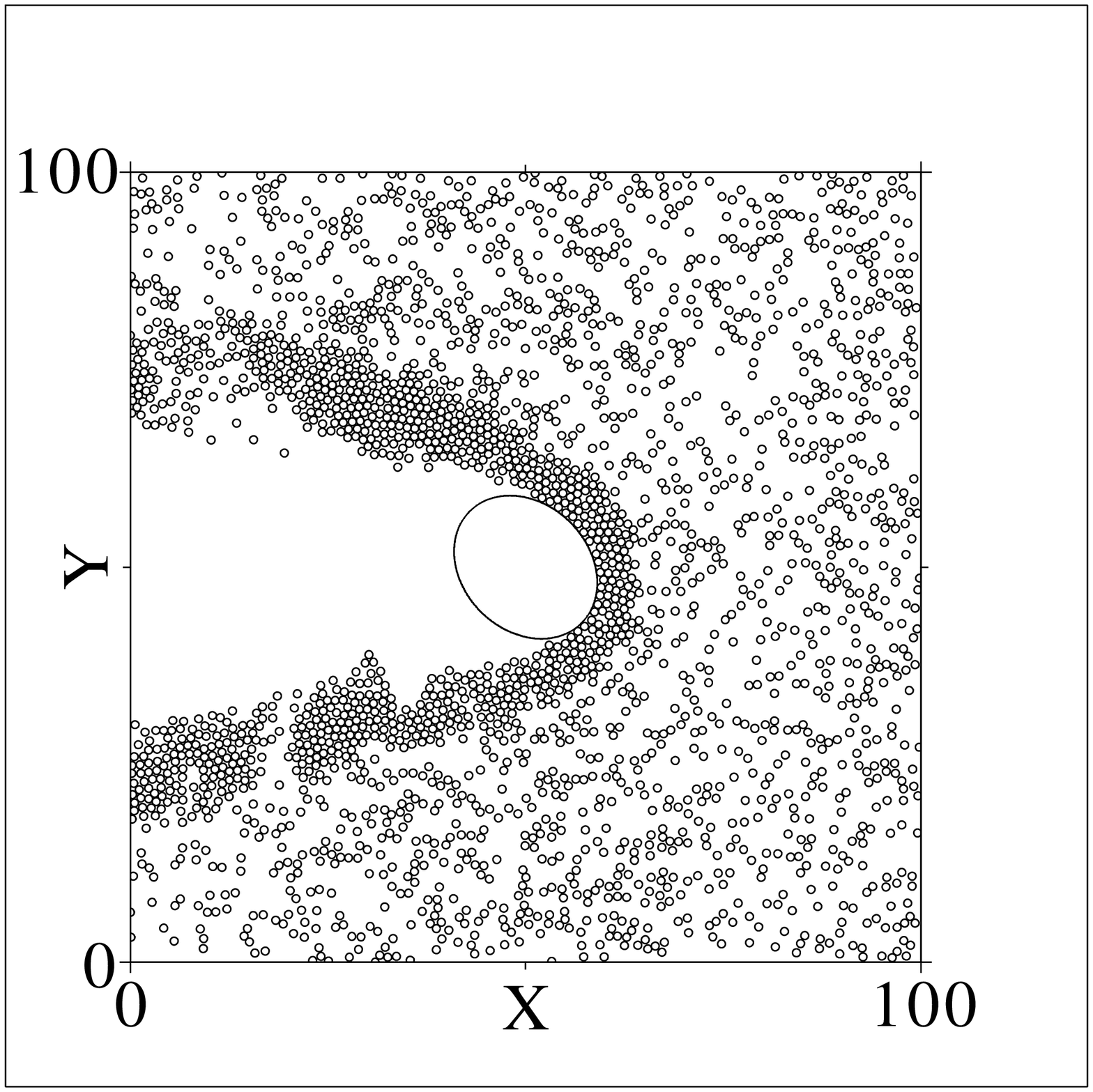,width=10.0cm,height=10.0cm}}\\
\rotatebox{0}{\epsfig{file=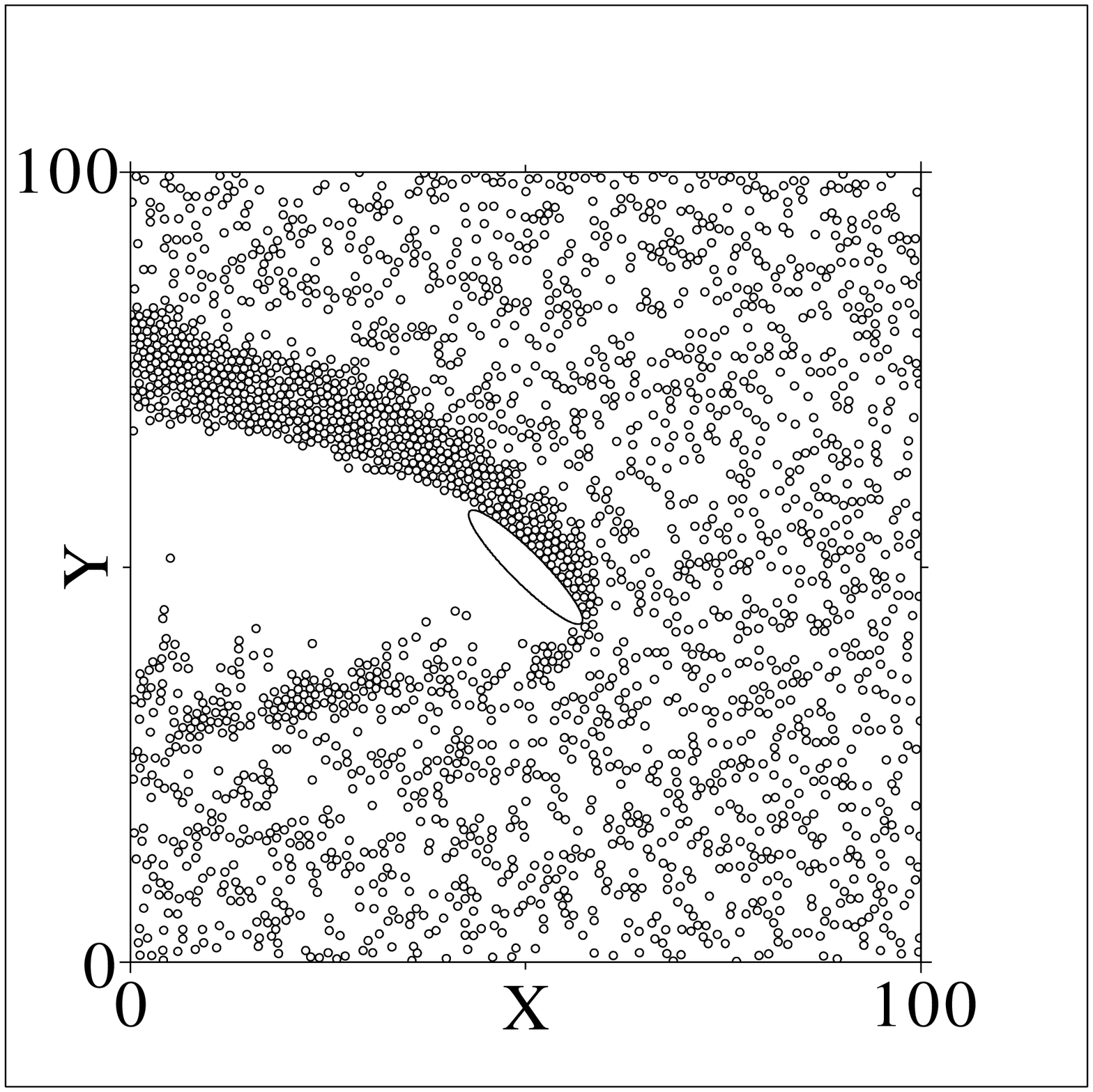,width=10.0cm,height=10.0cm}}\\
\end{tabular}
\caption{Two packing configurations for $a=10$, $\phi=-\pi/4$, $e=0.608$ and 
$k=0.80$ (top) and $k=0.20$ (bottom).\label{configs}}
\end{figure}

As a last note, the collision rate increases faster with $\phi$ 
for $k=0.20$ than for $k=0.80$. Fig. \ref{configs} has two configuration 
snapshots that illustrate the shock wave for two obstacles with distinct 
eccentricities.


\begin{figure}[h]
\rotatebox{0}{\epsfig{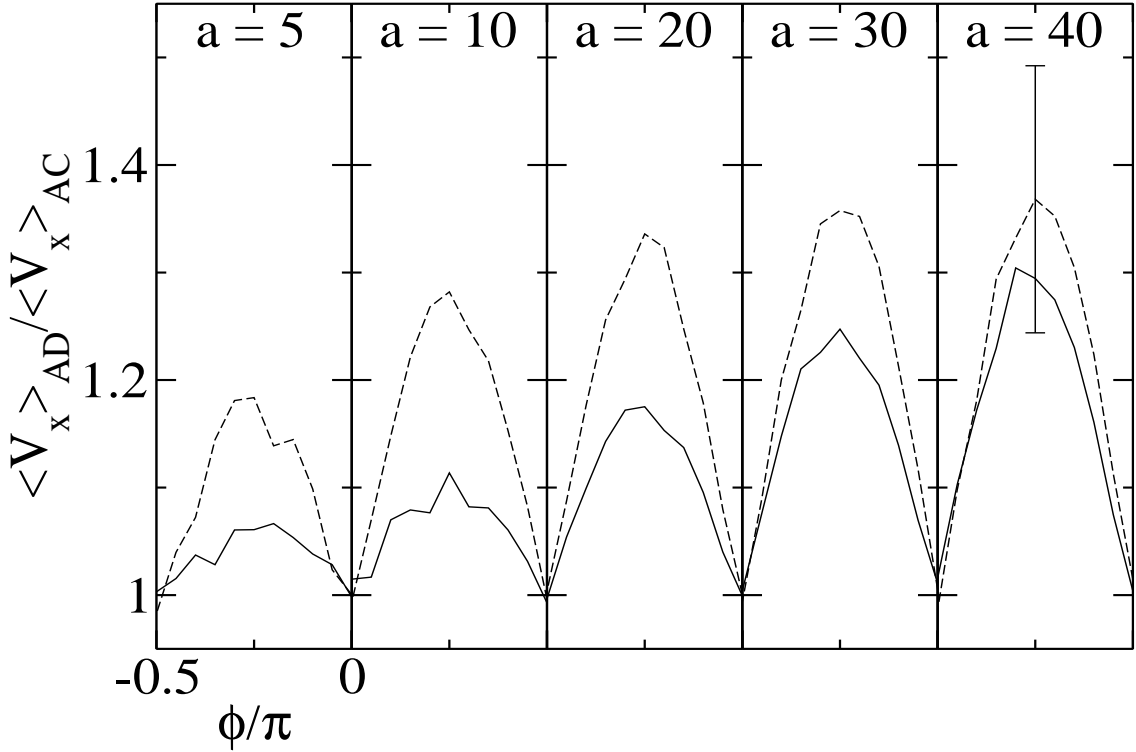}}
\caption{Ratio between the average horizontal velocity components in 
$\overline{AD}$ and $\overline{AC}$ for obstacles with $k=0.80$. Full lines 
correspond to packings with $e=0.952$, while dashed lines, to $e=0.608$. 
Scales in all plots are the same. 
\label{vx_ratio}}
\end{figure}
 
One can infer from these pictures the reason behind the asymmetrical peaks in 
the collision profiles when $\phi=-\pi/4$ and $k=0.20$. There is a very small 
concentration of particles in front of $\overline{AC}$, which certainly allows 
for more collisions to occur.

These considerations still left unanswered the question about the faster growth 
of $\left<L\right>$ with $a$ for stronger inelasticity compared to those at 
lower inelasticity. Since the net lift is the 
difference of the forces exerted in $\overline{AC}$ and $\overline{AD}$, one 
should seek a relative measure of the speeds in each of them in order to 
account for the size of the contributions each has in the net value. This is 
done by calculating the ratio between the average horizontal disk velocities in 
$\overline{AD}$ and $\overline{AC}$. If this ratio is greater than $1$, the 
collisions on $\overline{AD}$ exert, on average, a stronger force than those on 
$\overline{AC}$. These averages are evaluated directly from the simulations. 
The results, for all tilt angles and obstacle sizes, are 
given in fig. \ref{vx_ratio}. 

\begin{figure}[h]
\rotatebox{0}{\epsfig{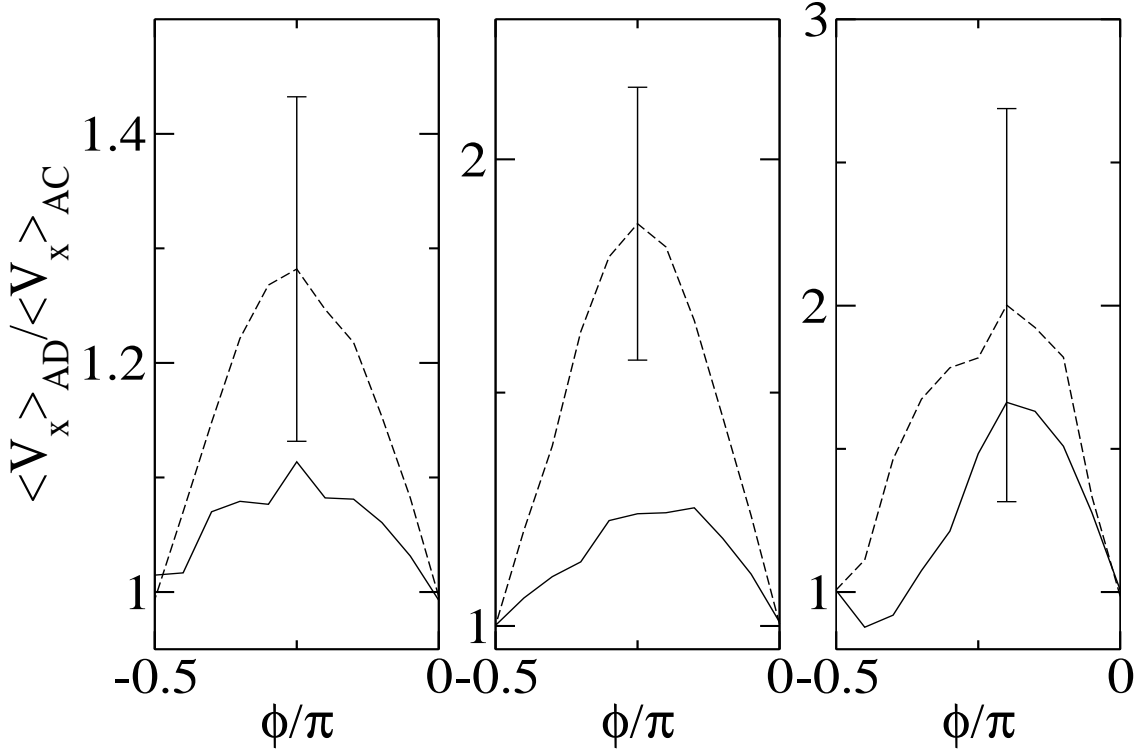}}
\caption{Ratio between the average horizontal velocity components in 
$\overline{AD}$ and $\overline{AC}$ for obstacles with $k=0.80$ (left panel), 
$k=0.50$ (middle panel) and $k=0.20$ (right panel). Full lines 
correspond to packings with $e=0.952$, while dashed lines, to $e=0.608$. 
\label{vx_ratio_k}}
\end{figure}

First it can be seen that except in a few cases, markedly at small obstacles, 
all ratios are larger than $1$, and those for $e=0.608$ cases are larger than 
those for $0.952$. It is possible to identify a trend in these graphs, despite 
the noise: the ratio grows with obstacle size. 
This is consistent with the results in fig. \ref{lift_phi}, where the net lift 
grows faster with $a$ for the $e=0.608$ flows compared to those with $e=0.952$. 
Fig. \ref{vx_ratio_k} has the data for the same quantity measured for obstacles 
with distinct eccentricities.




It is seen that the ratio increases as $k$ decreases (it reaches a factor of 
$2$ for $k=0.20$). These data, together with those for the collision profiles, 
show that the large lift observed for low $k$ obstacles is a consequence not 
only of a larger number of collisions at $\overline{AD}$, but also due to 
collisions with larger speeds than those at $\overline{AC}$.

\subsection{Oblique impacts\label{sub32}}
The last piece of this analysis is the role of the vertical velocity component 
in the lift value.
Before proceeding, the collision argument of section \ref{sec1} 
is reviewed by allowing the incoming disks to have a vertical velocity, $V$, 
and calculating the new momentum change in each collision. This will elucidate 
the effect of the vertical velocity on the lift value.

The velocity of a disk which collides with the obstacle is 
${\bf v}_0=-U{\bf i}+V{\bf j}$. The final disk velocity for a inelastic, 
frictionless collision is obtained as:
\begin{equation}
\label{new_vx}
v_x=U(e\cos^2\theta-\sin^2\theta)-\frac{1}{2}V(1+e)\sin2\theta,
\end{equation}
and
\begin{equation}
\label{new_vy}
v_y=\frac{1}{2}U(1+e)\sin2\theta+V(\cos^2\theta-e\sin^2\theta).
\end{equation}
The vertical momentum change due to this collision is given by:
\[
\Delta P_y=m(v_y-V),
\]
using the result for $v_y$, (\ref{new_vy}), and performing some algebra, the 
new momentum change is:
\begin{equation}
\label{new_P_change}
\Delta P_y=m(1+e)\left[\frac{1}{2}U\sin2\theta-V\sin^2\theta\right], 
\end{equation}
where $U>0$.

\begin{figure}[h]
\rotatebox{0}{\epsfig{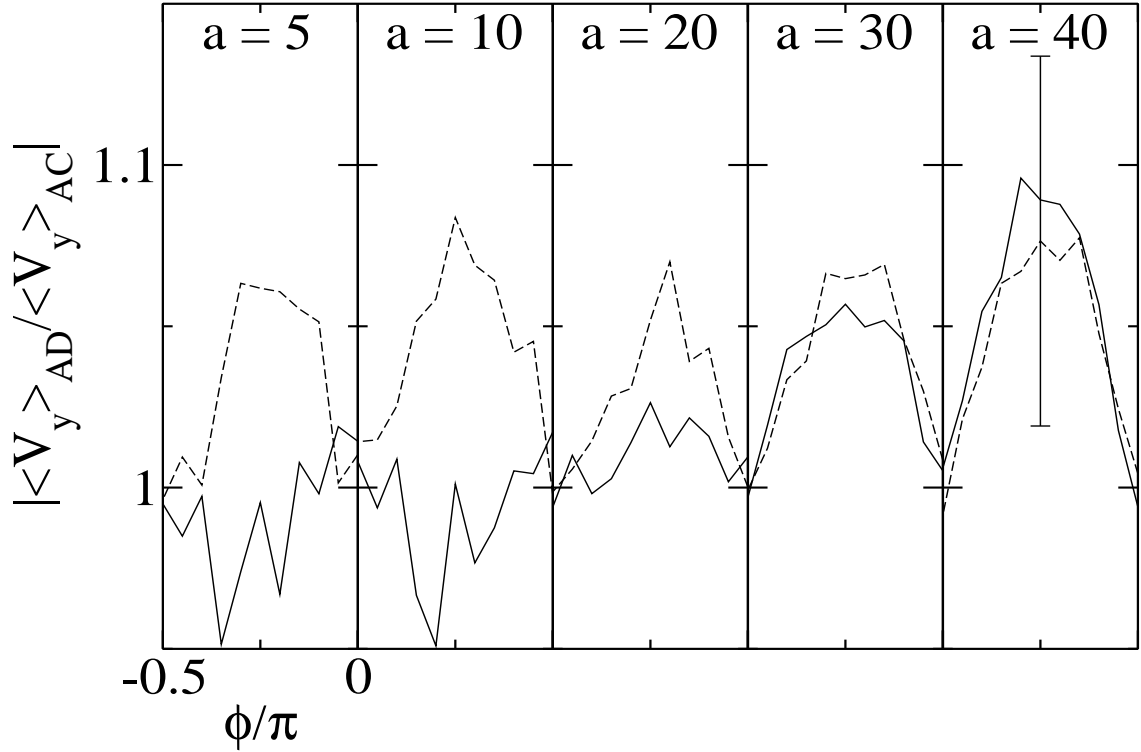}}
\caption{Ratio between the average vertical velocity components in 
$\overline{AD}$ and $\overline{AC}$. Full lines correspond to packings with 
$e=0.952$, while dashed lines, to $e=0.608$. Scales in all plots are 
the same. 
\label{vy_ratio}}
\end{figure}

\begin{figure}[h]
\rotatebox{0}{\epsfig{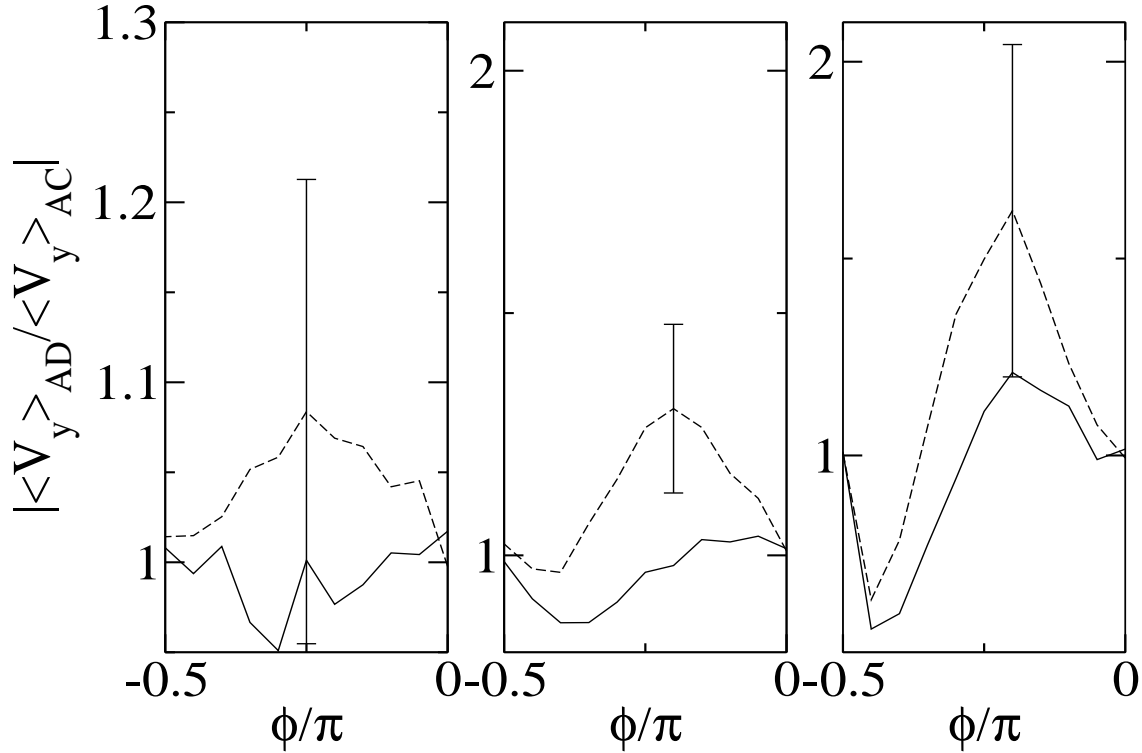}}
\caption{Ratio between the average vertical velocity components in 
$\overline{AD}$ and $\overline{AC}$. Full lines correspond to packings with 
$e=0.952$, while dashed lines, to $e=0.608$. The left panel is for $k=0.80$, 
the middle one, for $k=0.50$ and the right one, for $k=0.20$.
\label{vy_ratio_k}}
\end{figure}

By looking at eq. (\ref{new_P_change}) one can see that a collision that 
happens in the range $-\pi/2<\theta<0$, which corresponds to $\overline{AC}$, 
will produce a 
lower momentum change if $V<0$. Similarly, for collisions that take place in 
the range $0<\theta<\pi/2$, which is $\overline{AD}$, a lower momentum 
change will occur if $V>0$. From the velocity field data, fig. 
\ref{vel_fields}, it is clear that the average vertical velocities in 
$\overline{AD}$ and $\overline{AC}$ are positive and negative, respectively. 
The oblique impacts decrease the positive lift exerted by the flow 
in $\overline{AC}$, but also decrease the negative lift in $\overline{AD}$. 
Therefore, it is not obvious if both effects combined increase or decrease the 
net lift. In order for this combination of effects to increase the net lift, 
the vertical velocity in $\overline{AD}$ should be lower than the corresponding 
velocity in $\overline{AC}$. In figs. \ref{vy_ratio} and 
\ref{vy_ratio_k}, the 
absolute value of the ratio of the average vertical velocities in both segments 
is shown for distinct obstacle sizes eccentricities, respectively.


It is seen that the ratio, for $e=0.608$, depends little on the obstacle size 
and restitution coefficient, and that those for $e=0.952$ grow with $a$. It 
varies strongly with tile angle and eccentricity. Despite a few cases, the 
ratio is larger than $1$ which implies that the net lift 
decreases due to oblique impacts, as provided by the existence of the shock 
wave. 

\section{Conclusions\label{conclusions}}
A theoretical argument and numerical results on the net lift force exerted by 
a granular stream of equal disks on an ellipse were presented. 
The argument used to obtain a theoretical expression for the lift relied on 
inelastic, frictionless collisions of a very dilute granular flow, in the 
horizontal direction, and the ellipse. The expression obtained, eq. 
(\ref{lift_AD}), captures nicely the qualitative features of the numerical 
results. It does not, however, reproduce the results quantitatively, neither 
does it reproduce the difference in the lift force for $e=0.952$ and $e=0.608$.

The numerical data, shown in figs. \ref{lift_k} and \ref{lift_phi}, are lower 
than the theoretical prediction by a factor that depends mostly on the obstacle 
size (for $a=40$, eq. (\ref{lift_AD}) is about $4$ times larger than the 
measured results).

Additional analyses of the flow properties showed that, from the assumptions 
drawn in the text, the dilute flow one is the most problematic. The existence 
of the granular shock wave 
\cite{Amar01,Rericha02,Boudet08,Buchholtz98,Wass03} invalidates this 
hypothesis. This structure affects directly the velocity components of 
the disks at contact with the obstacle as well as the collision rate. The 
horizontal velocity at contact decreases due to dissipative collisions that 
occur before the disks reach the shock wave. Also, some of the horizontal 
momentum is deviated by the shock wave around the obstacle, introducing a 
vertical component whose net effect is to decrease the lift.

Perspectives to this work include a deeper analysis of dense packing effects 
on the lift. This would answer questions about the scaling of the lift with 
obstacle size, flow speed and density. Such analysis would also allows for a 
better understanding of the mechanism behind the appearance of the net lift, 
since an obvious consequence of the presence of the shock wave is that any 
collisions that occur on its edge should transmit some linear momentum through 
the dense region until pushing the disks closets to the obstacle, and that is 
when the actual force is made. Force transmission models, such as those in 
\cite{Liu95,Ostojic05} can be good starting points for this analysis. 
Moreover, the effect of eccentricity on 
the shock wave is something worth investigating since, as seen here, for a 
constant flow speed, there should be some eccentricity value that allows for 
the appearance of the gap between the shock wave and the obstacle.
Finally, the hydrodynamical approach to granular 
flow could also be tested in such a situation. As argued in the text, ignoring 
the disk size does not affect the results for large obstacles. Hence, if the 
obstacle is much larger than the disks, the predictions of such theory should 
be realized in simulations, since scale separation could be achieved, at least 
approximately.

\section*{Acknowledgements} 
I thank A. P. F. Atman and J. C. Costa for a critical reading of this 
manuscript. This work is financially supported by CNPq and FAPESPA.


\begin{thebibliography}{100}
\bibitem{Jaeger96} H. M. Jaeger, S. R. Nagel, R. P. Behringer, Rev. Mod. Phys. 
{\bf68}, 1259 (1996).
\bibitem{Wieghardt75} K. Wieghardt, Annu. Rev. Fluid Mech. {\bf 7}, 89 (1975).
\bibitem{Campbell90} C. S. Campbell, Annu. Rev. Fluid Mech. {\bf 22}, 57 
(1990).
\bibitem{Gold03} I. Goldhirsch, Annu. Rev. Fluid Mech. {\bf 35}, 267 (2003).
\bibitem{Midi04} GDR MiDi, Eur. Phys. J. E {\bf 14}, 41 (2004).
\bibitem{Amar01} Y. Amarouchene, J. F. Boudet, H. Kellay, Phys. Rev. Lett. 
{\bf 86}, 4286 (2001).
\bibitem{Rericha02} E. C. Rericha, C. Bizon, M. D. Shattuck, H. L. Swinney, 
Phys. Rev. Lett. {\bf 88}, 014302 (2002).
\bibitem{Boudet08} J. F. Boudet, Y. Amarouchene, H. Kellay, Phys. Rev. Lett. {\bf 101}, 254503 (2008).
\bibitem{Albert99} R. Albert, M. A. Pfeifer, A.-L. Barabasi, P. Schiffer, Phys. Rev. Lett. {\bf 82}, 205 (1999).
\bibitem{Chehata03} D. Chehata, R. Zenit, C. R. Wassgren, Phys. Fluids {\bf 15}, 1622 (2003).
\bibitem{Buchholtz98} V. Buchholtz, T. Poeschel, Granular Matter {\bf 1}, 33 
(1998).
\bibitem{Wass03} C. R. Wassgren, J. A. Cordova, R. Zenit, A. Karion, Phys. Fluids {\bf 15}, 3318 (2003).
\bibitem{Ciamarra04} Pica Ciamarra M. {\em et al}., Phys. Rev. Lett. {\bf 92}, 194301 (2004).
\bibitem{Soller06} R. Soller, S. A. Koehler, Phys. Rev. E {\bf 74}, 021305 
(2006).
\bibitem{Ding10} Y. Ding, N. Gravish, D. I. Goldman, Phys. Rev. Lett. {\bf 106}, 028001 (2011).
\bibitem{rapaport} D. C. Rapaport, 
{\em The art of molecular dynamics simulation}, 2nd ed., Cambridge University Press (Cambridge, 2007).
\bibitem{Silbert01} L. E. Silbert {\em et al}., Phys. Rev. E {\bf 64}, 051302 (2001).
\bibitem{Dziu01} A. Dziugys, B. Peters, Int. J. Numer. Anal. Meth. Geomech. 
{\bf 25}, 1487 (2001).
\bibitem{nr} W. H. Press, B. P. Flannery, S. A. Teukolsky, W. T. Vetterling, 
{\em Numerical recipes in C}, 2nd edition, Cambridge University Press (Cambridge, 1986).
\bibitem{Deng10} Y. H. Deng, J. J. Wylie, Q. Zhang, Phys. Rev. E {\bf 82}, 
011307 (2010).
\bibitem{Liu95} C.-h. Liu {\em et al}., Science {\bf 269}, 513 (1995).
\bibitem{Ostojic05} S. Ostojic, D. Panja, Europhys. Lett. {\bf 71}, 70 (2005).
\end{thebibliography}

\end{document}